\documentclass[runningheads]{svmult}
\usepackage{makeidx}   
\usepackage{graphicx}  
\usepackage{subeqnar}  
\usepackage{multicol}  
\usepackage{cropmark} 
\usepackage{physprbb}  
\newcommand{\gsim}{\mathrel{\mathop{\kern 0pt \rlap
  {\raise.2ex\hbox{$>$}}}
  \lower.9ex\hbox{\kern-.190em $\sim$}}}
\newcommand{\lsim}{\mathrel{\mathop{\kern 0pt \rlap
  {\raise.2ex\hbox{$<$}}}
  \lower.9ex\hbox{\kern-.190em $\sim$}}}
\def\Journal#1#2#3#4{{#1} {#2} (#4) {#3}}

\def\APP{\em Astrop. Phys.}

\def\NCC{{\em Il Nuovo Cimento} {\bf C}}

\def\NJP{\em New Journal of Physics}
\def\NPA{{\em Nucl. Phys.} {\bf A}}

\def\PLB{{\em Phys. Lett.}  {\bf B}}

\def\PRD{{\em Phys. Rev.} {\bf D}}

\begin{document}
\title*{\footnotesize ROM2F/2002/12 (May 2002) 
to appear in the Proceed. of DARK2002.
\protect\newline 
\Large \protect\newline WIMP Search by DAMA at Gran
Sasso}
\toctitle{WIMP Search by DAMA at Gran Sasso}
%
%
\titlerunning{WIMP Search by DAMA at Gran Sasso}
%
\author{R. Bernabei\inst{1}
\and M. Amato\inst{2}
\and P. Belli\inst{1}
\and F. Cappella\inst{1}
\and R. Cerulli\inst{1}
\and C.J. Dai\inst{3}
\and H.L. He\inst{3}
\and G. Ignesti\inst{2}
\and A. Incicchitti\inst{2}
\and H.H. Kuang\inst{3}
\and J.M. Ma\inst{3}
\and F. Montecchia\inst{1}
\and F. Nozzoli\inst{1}
\and D. Prosperi\inst{2}}
\authorrunning{R. Bernabei et al.}
%
%
%
\institute{Dip.to di Fisica, Universit\`a di Roma ``Tor Vergata''
        and INFN, sez. Roma2, I-00133 Rome, Italy
\and Dip.to di Fisica, Universit\`a di Roma ``La Sapienza''
        and INFN, sez. Roma, I-00185 Rome, Italy
\and    IHEP, Chinese Academy,  P.O. Box 918/3,
        Beijing 100039, China.}
\maketitle
\begin{abstract}
DAMA is searching for rare processes by developing and using
several kinds of radiopure scintillators: in particular, NaI(Tl),
liquid Xenon and CaF$_2$(Eu). Here only the results released so far
on the WIMP annual modulation signature are summarized and compared
with
results from other experiments, including the recent re-analysis of
CDMS-I data. Next perspectives are also shortly addressed.
\end{abstract}
%

\section{Introduction}

DAMA is devoted to the search for rare processes
(such as WIMPs direct detection,
$\beta\beta$ processes, charge-non-conserving processes,
Pauli exclusion principle violating
processes, nucleon instability, solar
axions and exotics
\cite{tutti,Benc,Bepsd,Mod1,Beapp,Mod2,Ext,Mod3,Sist,Sisd,Inel,Hep})
by developing and using low radioactive scintillators.
The main experimental set-ups, which are running at present, are: the
$\simeq$ 100 kg NaI(Tl) set-up, the $\simeq$ 6.5 kg liquid Xenon (LXe)
set-up and the so-called ``R\&D'' apparatus.
Moreover, a low-background germanium detector is operative underground
since many years for measurements and selections of samples.

In this paper only the results obtained in the search for WIMPs
by exploiting the annual modulation signature with
the
$\simeq$ 100 kg NaI(Tl) set-up, their comparison
with those of other
experiments (including the more recent re-analysis of the CDMS-I data)
and some perspectives of the new LIBRA (Large sodium
Iodine Bulk for RAre processes) set-up in preparation
will be addressed.

A full description of the $\simeq$ 100 kg NaI(Tl) set-up and of
its performances can be found in
ref.\cite{Beapp}. It is worth to note that some upgrading has been
performed since then;
in particular, during August 2000 the electronic chain and the DAQ
have been fully substituted achieving improved performances.

The recoil/electron light ratio for $^{23}$Na and $^{127}$I
and the pulse shape discrimination
capability have been measured by neutron source and upper
limits on recoils have been
measured in this set-up by exploiting
the pulse shape discrimination technique
\cite{Bepsd}. Moreover,
studies on possible
diurnal variation of the low energy rate in the data of the $\simeq$
100 kg NaI(Tl) set-up have also been carried out.
It could be expected because of the Earth's daily rotation;
in fact, during the sidereal day the Earth shields a given detector
with a variable thickness, eclipsing the ``wind'' of Dark Matter
particles
but only in case of high cross section candidates (to which small
halo fraction would correspond). By analyzing a statistics of
14962 $kg \times day$
no evidence for diurnal rate variation with sidereal time has been
observed \cite{Benc};
this result supports that the effect pointed out
by the studies on the WIMP annual modulation signature (see later)
would account for a halo fraction $\gsim 10^{-3}$ \cite{Benc}.

The main goal of the $\simeq$
100 kg NaI(Tl) set-up is to investigate
the WIMP annual modulation signature.
In fact, the WIMPs are embedded in the galactic halo; thus,
our solar system, which is moving with respect to the galactic system,
is continuously hit by a WIMP ``wind'' which can be mainly searched for
by WIMP elastic scattering on the target nuclei of the detector.
In particular,
since the Earth rotates around the Sun,
which is moving with respect to the galactic system,
it would be crossed by a larger WIMP flux in June (when its
rotational velocity is summed to
the one of the solar system with respect to the Galaxy)
and by a smaller one in December (when the two velocities are subtracted).
The fractional difference between the maximum and the minimum of the rate
is expected to be of order of $\simeq$ 7\%.

\vskip -0.05cm

The $\simeq$ 100 kg highly radiopure NaI(Tl) DAMA set-up \cite{Beapp}
can effectively exploit such a signature because of its
well known technology, of its high intrinsic radiopurity, of its mass,
of its suitable control of all the operational parameters and
of the deep underground experimental site.

\vskip -0.05cm

The annual modulation signature is very distinctive as we have
already pointed out \cite{Mod1,Beapp,Mod2,Ext,Mod3,Sist,Sisd,Inel}.
In fact, a WIMP-induced seasonal effect must simultaneously satisfy
all the following requirements: the rate must contain a component
modulated according to a cosine function (1) with one year period (2)
and a phase that peaks around $\simeq$ 2$^{nd}$ June (3);
this modulation must be found
in a well-defined low energy range, where WIMP induced recoils
can be present  (4); it must apply to those events in
which just one detector of many actually "fires", since
the WIMP multi-scattering probability is negligible (5); the modulation
amplitude in the region of maximal sensitivity must be $\lsim$7$\%$ (6).
Only systematic effects able
to fulfil these 6 requirements could fake this signature;
no one able to do that has been found or suggested \cite{Sist}, on the
contrary of
what sometimes claimed by some author (see eg. ref. \cite{Spoo}).

\vskip -0.05cm

Results obtained by investigating the annual modulation
signature in the data collected during the first four annual
cycles have been released so far \cite{Mod1,Mod2,Ext,Mod3,Sist,Sisd,Inel}.
The latter ones
are summarized below and compared with those of some other experiments.
At present the experiment is running taking data for the 7$^{th}$
cycle; at the end of this cycle the new LIBRA set-up ($\simeq$ 250
kg of NaI(Tl)) will be installed.

\vskip -0.4cm

\section{A Model Independent Analysis of the
\protect\newline Annual Modulation Data}

The annual modulation signature
offers the possibility to obtain a model independent evidence for the
presence of a WIMP component in the galactic halo. The large number of
peculiarities (see above), which have to be satisfied for that, assures
the absence of possible known systematic effects able to mimic such a
signature (see above and ref. \cite{Sist}) and, therefore, an
effective tool of investigation, as originally pointed out in ref.
\cite{Free}.

\begin{figure}[htb]
\centering
\vskip -1.2cm
\includegraphics[height=5.0cm]{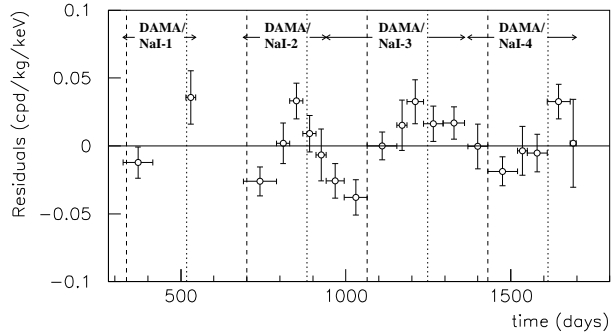}
\caption{Model independent residual rate for single hit events, in the
2--6 keV cumulative energy interval, as a function of the time elapsed
since
January 1-st of the first year of data taking. The expected
behaviour of a WIMP signal is
a cosine function with minimum roughly at the dashed
vertical lines and with maximum roughly at the dotted ones.}
\label{fg:fig1}
\vskip -0.6cm
\end{figure}

The data released so far refer to a total statistics of 57986 $kg \cdot
day$ collected during four independent experiments of one year cycle each
one.
A model independent analysis of these data
offers an immediate evidence of the presence of
an annual modulation of the rate of the single hit events
in the lowest energy interval (2 -- 6 keV) as shown in Fig. \ref{fg:fig1}.
There each data point has been obtained
from the raw rate measured in the corresponding time interval,
after subtracting the constant part.
The results of the four years give
consistent results with proper period and phase.
In particular, the $\chi^2$ test on the data of Fig.
\ref{fg:fig1}
disfavors the hypothesis of unmodulated behaviour
(probability: $4 \cdot 10^{-4}$).
No modulation is found in higher energy regions.

We have extensively discussed the results of the investigations
of all the possible known systematics when releasing
the data of each annual cycle; moreover, a dedicated
paper \cite{Sist} has been released on this topic.
No known systematic effect or side reaction able to mimic a WIMP induced
effect has been found as discussed in details in ref. \cite{Sist}.
In particular, no effect mentioned in ref. \cite{Spoo} can be able to
mimic it (see e.g. ref. \cite{Sist}).

In conclusion, a WIMP contribution
to the measured rate is candidate by the result of the model independent
approach independently on the nature and coupling
with ordinary matter of the possible WIMP particle.

\section{Model Dependent Analyses of the
\protect\newline Annual Modulation Data}

To investigate the nature and coupling
with ordinary matter of a possible candidate, a suitable energy and time
correlation analysis is necessary as well as a complete model
framework. We remark that a model framework is identified
not only by general
astrophysical, nuclear and particle physics assumptions, but also by the
set
of values used for all the parameters needed in the model itself and
in related quantities
(for example WIMP local velocity, $v_0$, form factor parameters, etc.).

For simplicity, initially we have considered the case of purely
spin-independent coupled WIMP. In fact,
often the spin-independent interaction
with ordinary matter is assumed to be dominant since e.g.
most of the used target-nuclei are practically
not sensitive to SD interactions as on the contrary
$^{23}$Na and $^{127}$I are and the theoretical calculations are even
more complex.

Moreover, the simplest model scenario has been considered as
well as fixed parameters values \cite{Mod1,Mod2}. Then, this case has
been extended by considering the many uncertainties which exist
on the astrophysical velocity distribution \cite{Ext,Mod3} and the
physical constraint
which arises from the measured upper limits on recoils \cite{Mod3}. Then,
some of the other possible particle scenarios
have been considered such as extensions to the general case of WIMPs with
both spin-independent (SI) and spin-dependent (SD) coupling \cite{Sisd}
and to that of WIMPs with inelastic scattering \cite{Inel}.
In these latter cases, the effect of the uncertainties on some other
parameters has been included. Moreover,
recently an investigation on the effect induced by different
consistent halo models on the result for purely SI coupled WIMPs has also
been carried out in ref. \cite{Hep}.

At present the lightest supersymmetric
particle named neutralino is considered the best candidate for WIMP.
Note, in particular, that the
results of the data analyses \cite{Mod3,Sisd,Inel}
summarized here and in the
following hold for the neutralino, but are not restricted only to
this candidate.

In supersymmetric theories
both the squark and the Higgs bosons exchanges give contribution to
the coherent (SI) part of the neutralino cross section, while
the squark and the $Z^{0}$ exchanges give contribution
to the spin dependent (SD) one. Therefore,
 the differential energy distribution of the recoil nuclei
in WIMP-nucleus elastic scattering
can be calculated \cite{Bepsd,Botdm}
by means of the differential cross section of the WIMP-nucleus elastic
processes: $\frac{d\sigma}{dE_R}(v,E_R) = \left( \frac{d \sigma}{dE_{R}}
\right)_{SI}+
\left( \frac{d \sigma}{dE_{R}} \right)_{SD}$
where $v$ is the WIMP velocity in the laboratory frame and
$E_R$ is the recoil energy.

\subsection
{WIMPs With Dominant SI Interaction in
\protect\newline a Given Model Framework}

\normalsize

\begin{figure}[!h]
\centering
\includegraphics[angle=90,height=4.3cm]{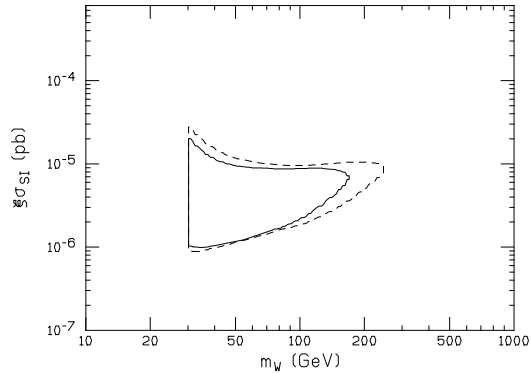}
\vskip 0.2cm
\caption{A purely SI case: regions allowed at 3$\sigma$ C.L. on the plane
$\xi\sigma_{SI}$ versus m$_W$
for a WIMP with dominant SI interaction and mass above 30 GeV in the
model framework considered in ref. \cite{Mod3}:
i) when $v_0$ uncertainty (170 km/s $\le v_0 \le 270$ km/s;
continuous contour) has been included; ii) when
also a possible bulk halo rotation as in ref. \cite{Ext} (dashed contour)
is
considered. Note that these regions hold for the given model framework
(assumptions and parameters) and correspond to the superposition of all
the
3 $\sigma$ allowed regions obtained when varying $v_0$ in the allowed
range; thus many "most likely" values correspond to it.
Moreover, as widely known,
the inclusion of present uncertainties on some other astrophysical,
nuclear and particle
physics parameters would enlarge these regions (varying again
consequently the
"most likely" values for each
considered set).
Some partial discussions can be found in \cite{Ext,Mod3,Hep}.}
\label{fg:fig2}
\end{figure}

\begin{figure}[htb]
\centering
\vskip -1.7cm
\includegraphics[height=7.cm]{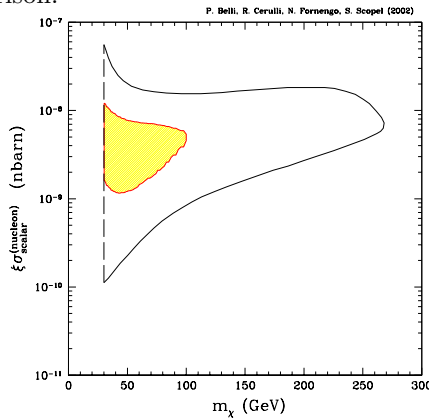}
\vskip -0.6cm
\caption{Effect of halo modelling:
the external continuous contour is the 3$\sigma$ annual modulation region
for purely SI coupled WIMPs in the plane $\xi \sigma_{SI}$ (where
$\sigma_{SI} = \sigma_{scalar}^{(nucleon)}$)
versus WIMP mass, obtained by considering all the (non-rotating) galactic
halo models discussed in ref. \cite{Hep}. To point out the effect of the
halo modelling alone, there is also shown the internal dashed region
which corresponds to a simplified model of isothermal galactic
halo in the same framework as ref. \cite{Mod3}, but
assuming here the $v_0$ value at 220 km/s.}
\label{fg:fig3}
\vskip -0.4cm
\end{figure}

As first scenario a full energy and time correlation
analysis -- properly accounting for the physical constraint arising
from the measured upper limit on recoils \cite{Bepsd,Sist}
-- has been carried out in the framework of a given
model for purely spin-independent coupled candidates with mass above 30
GeV.
A standard
maximum likelihood method has been used.
Following the usual procedure we have built the $y$ log-likelihood
function,
which depends on the experimental data and on the
theoretical expectations for the considered model framework;
then, $y$ is minimized and parameters' regions allowed
at given confidence level are derived.
Note that different model frameworks (see above)
vary the expectations and, therefore, the cross section and mass values
corresponding to the $y$ minimum, that is also the allowed region
at given C.L.. In particular,
the inclusion of the uncertainties associated to the models and to
every parameter in the models themselves
as well as other possible scenarios largely
enlarges the allowed region
as discussed e.g. in ref. \cite{Ext}
for the particular case of the astrophysical velocities.
In the case considered in ref. \cite{Mod3} the minimization procedure
has been repeated by varying the WIMP local velocity, $v_0$, from 170
km/s to 270 km/s
to account for its present uncertainty.
For example, the values
$m_W = (72^{+18}_{-15})$ GeV and
$\xi \sigma_{SI} = (5.7 \pm 1.1) \cdot 10^{-6}$ pb
correspond to the position of $y$ minimum when $v_0$ = 170 km/s,
while $m_W = (43^{+12}_{-9})$ GeV and
$\xi \sigma_{SI} = (5.4 \pm 1.0) \cdot 10^{-6}$ pb
when $v_0$ = 220 km/s. Here, $\xi$ is the WIMP local density
in 0.3 GeV cm$^{-3}$ unit,
$\sigma_{SI}$ is the point-like SI WIMP-nucleon
generalized cross section and $m_W$ is the WIMP mass.
Fig. \ref{fg:fig2} shows the regions allowed at 3$\sigma$
C.L. in such a model framework,
when the uncertainty on $v_0$ is taken into account (solid contour) and
when possible bulk halo rotation is considered (dashed contour).
For simplicity, no other uncertainty on the used parameters has been
considered there
(some of them have been included in the approach summarized in the next
subsection \cite{Sisd}); thus, obviously the real allowed region
including the effect of all the uncertainties is much larger
than quoted for this
simplified model framework as discussed at some extent also in ref.
\cite{Mod3} and recently shown -- as regards the halo model effect -- in
ref. \cite{Hep} (see Fig. \ref{fg:fig3}).

A quantitative comparison between the residuals and the modulation
amplitudes for this particular model framework
has been discussed
in ref. \cite{Mod3}. Moreover, we remark that the energy and time
behaviour of the modulation amplitudes and residuals as well as the
accounting for known uncertainties (such as quenching factors, form
factors parameters, etc.) play a crucial role to obtain a
meaningful comparison.

In conclusion, the observed effect investigated in terms of
a WIMP candidate with dominant SI interaction and mass above 30 GeV
in the simplified model framework considered in ref. \cite{Mod3},
supports allowed WIMP masses up to about 105 GeV (1 $\sigma$ C.L.)
and even
up to about 250 GeV (1 $\sigma$ C.L.) if different halo models
are considered \cite{Hep}.
Lower $\xi\sigma_{SI}$ would be implied by the inclusion of known
uncertainties on parameters (for example, on the quenching factors and
on the
form factor parameters)
and on model features.

Theoretical implications of these results in terms
of a neutralino with dominant SI interaction and mass above 30 GeV
have been discussed in ref. \cite{Botdm,Arn},
while the case for an heavy neutrino of the fourth family
has been considered in ref. \cite{Far}.
Let us comment that a correct evaluation and interpretation of the
theoretical results are necessary; in fact,
sometimes the uncertainties on the local halo density are not taken
into account and the rescaling procedures are not often
applied to correctly evaluate the WIMP local density.
They must be considered for a reliable and correct presentation
as well as the uncertainties which exist on some aspects of
supersymmetric models
in order to avoid that the given figure "would drive"
the reader to wrong conclusion.
Finally, note that the density of points in the calculated scatter plots
do not represent
a probability density.

\subsection
{ WIMPs With Mixed Coupling in Given Model Framework}

\normalsize

Since the $^{23}$Na and $^{127}$I nuclei are sensitive
to both SI and SD couplings -- on the contrary e.g. of
$^{nat}$Ge and $^{nat}$Si which are sensitive mainly
to WIMPs with SI coupling (only 7.8 \% is non-zero spin isotope in
$^{nat}$Ge
and only 4.7\% of $^{29}$Si in $^{nat}$Si) --
the analysis of the data has been extended
considering the more general case \cite{Sisd}.
This implies a WIMP having not only a spin-independent,
but also a spin-dependent coupling different from zero, as it is also
possible e.g. for the neutralino.

\begin{figure}[htb]
\centering
\includegraphics[height=8.cm]{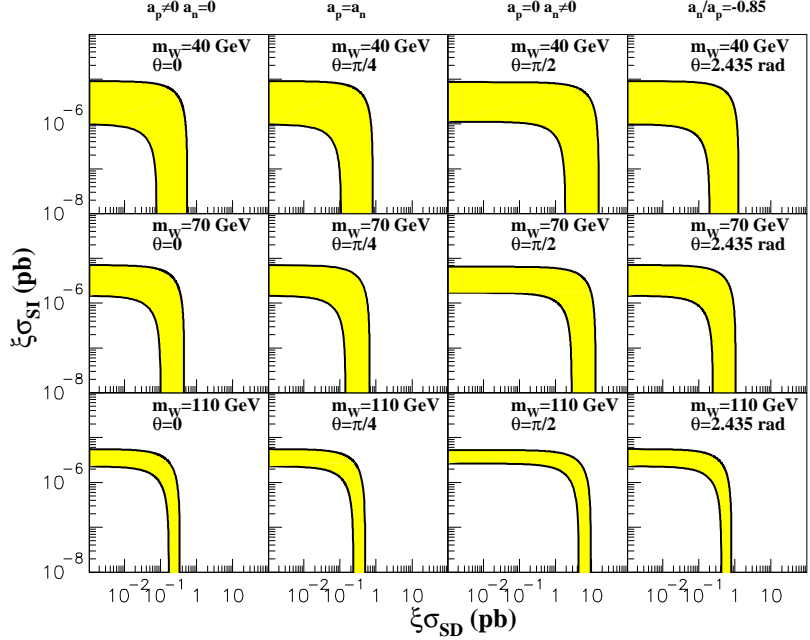}
\caption{A mixed SI/SD case: example of slices
of the region allowed at 3 $\sigma$ C.L.
in the ($\xi \sigma_{SI}$, $\xi \sigma_{SD}$, $m_W$) space for some
$m_W$ and $\theta$
values in the model framework considered in ref.~\cite{Sisd}.
Only four particular couplings are reported here for simplicity:
i) $\theta$ = 0;
ii) $\theta$ = $\pi/4$
iii) $\theta$ = $\pi/2$; iv)
$\theta$ = 2.435 rad. Note that e.g. Ge experiments are sensitive mainly
only
to SI coupling and, therefore, cannot explore most of the DAMA allowed
regions in
this scenario.}
\label{fg:fig4}
\vskip -0.4cm
\end{figure}

Then, the log-likelihood function has been minimized -- properly
accounting also for the physical
constraint set by the measured upper limit on recoils \cite{Bepsd} --
and parameters' regions allowed
at given confidence level have been obtained.
In particular, the calculation has been performed
by minimizing the $y$ function with respect to
the $\xi \sigma_{SI}$, $\xi \sigma_{SD}$ and $m_W$
parameters for each given $\theta$ value.
Here,  $\sigma_{SD}$  is the point-like SD WIMP
cross section on nucleon and $tg\theta$ is the ratio
between the effective SD coupling constants on neutrons, $a_n$,
and on proton, $a_p$; therefore, $\theta$ can assume values between
0 and $\pi$ depending on the SD coupling.
In the present framework the uncertainties on $v_0$ have been included;
moreover, the uncertainties on the nuclear
radius and the nuclear surface thickness parameter in the SI form factor,
on
the $b$ parameter in the used SD form factor
and on the measured quenching factors \cite{Bepsd} of these detectors
have also been considered \cite{Sisd}.
For simplicity, Fig. \ref{fg:fig4} shows slices for some
$m_W$ of the region allowed at 3 $\sigma$ C.L.
in the ($\xi \sigma_{SI}$, $\xi \sigma_{SD}$, $m_W$) space
for four particular couplings:
i)  $\theta$ = 0 ($a_n$ =0 and $a_p \ne$ 0 or $|a_p| >> |a_n|$);
ii) $\theta = \pi/4$ ($a_p = a_n$);
iii)  $\theta$ = $\pi/2$ ($a_n \ne$ 0 and $a_p$ = 0 or  $|a_n| >> |a_p|$);
iv) $\theta$ = 2.435 rad ($ \frac {a_n} {a_p}$ = -0.85, pure Z$^0$
coupling).
The case $a_p = - a_n$ is nearly similar to the case iv).

As already pointed out, when the SD contribution goes to zero
(y axis in Fig. \ref{fg:fig4}), an interval not
compatible with zero is obtained for $\xi\sigma_{SI}$.
Similarly, when the SI contribution goes to zero
(x axis in Fig. \ref{fg:fig4}), finite values for the SD cross section are
obtained.
Large regions are allowed for mixed configurations
also for $\xi\sigma_{SI} \lsim 10^{-5}$ pb and
$\xi\sigma_{SD} \lsim 1$ pb; only in the particular case of
$\theta = \frac {\pi} {2}$ (that is $a_p = 0$ and $a_n \ne 0$)
$\xi\sigma_{SD}$ can increase up to $\simeq$ 10 pb, since
the $^{23}$Na and $^{127}$I nuclei have the proton as odd
nucleon. Moreover, in ref. \cite{Sisd} we have also pointed out that:
i) finite values can be allowed for $\xi\sigma_{SD}$
even when $\xi\sigma_{SI} \simeq 3 \cdot 10^{-6}$ pb as in the region
allowed
in the pure SI scenario considered in the previous subsection;
ii) regions not compatible with zero
in the $\xi\sigma_{SD}$ versus m$_W$ plane are allowed
even when $\xi\sigma_{SI}$ values much lower than those allowed in
the dominant SI scenario previously summarized are considered; iii)
minima of the $y$ function with both $\xi\sigma_{SI}$ and
$\xi\sigma_{SD}$ different from zero are present for some $m_W$ and
$\theta$ pairs; the related confidence level ranges between
$\simeq$ 3 $\sigma$ and $\simeq$ 4 $\sigma$ \cite{Sisd}.

Further investigations are in progress on these
model dependent analyses to account for other known parameters
uncertainties
and for possible different model assumptions.
As an example we recall
that for the SD form factor
an universal formulation is not possible since
the internal degrees of the WIMP particle model (e.g. supersymmetry
in case of neutralino) cannot be completely separated
from the nuclear ones. In the calculations presented here we have adopted
the SD form factors of ref. \cite{Ress97} estimated by considering
the Nijmengen nucleon-nucleon potential. Other formulations are
possible for SD form factors and can be considered with evident
implications
on the obtained allowed regions.

In conclusion, this analysis has shown that the DAMA data
of the four annual cycles, analysed in terms
of WIMP annual modulation signature,
can also be compatible with a mixed scenario
where both $\xi\sigma_{SI}$ and $\xi\sigma_{SD}$ are
different from zero.

\subsection
{Inelastic Dark Matter}
\normalsize

It has been suggested \cite{Wei01} that the
observed annual modulation effect could be induced by possible
inelastic Dark Matter: relic particles that prefer to scatter
inelastically
off of nuclei. The inelastic Dark Matter
could arise from a massive complex scalar split into two approximately
degenerate real scalars or from a Dirac fermion split into two
approximately degenerate Majorana fermions, namely $\chi_+$ and $\chi_-$,
with a $\delta$ mass splitting. In particular, a specific
model featuring a real component of the sneutrino,
in which the mass splitting naturally arises, has been given in ref.
\cite{Wei01}.
It has been shown that for the $\chi_-$ inelastic scattering
on target nuclei a kinematical constraint exists which favours
heavy nuclei (such as $^{127}$I) with respect to
lighter ones (such as e.g. $^{nat}$Ge) as target-detectors media.
In fact, $\chi_{-}$ can only inelastically scatter
by transitioning to $\chi_{+}$ (slightly heavier state than $\chi_{-}$)
and this process can occur
only if the $\chi_{-}$ velocity is larger than
$v_{thr} = \sqrt{\frac{2\delta}{m_{WN}}}$
where $m_{WN}$ is the WIMP-nucleus reduced mass ($c=1$).
This kinematical constraint becomes increasingly severe
as the nucleus mass, $m_N$, is decreased \cite{Wei01}.
Moreover, this model scenario gives rise -- with respect to the
case of WIMP elastically scattering -- to an enhanced
modulated component, $S_m$, with respect to the unmodulated one, $S_0$,
and to largely different behaviours with energy for
both $S_0$ and $S_m$ (both show a higher mean value) \cite{Wei01}.

A dedicated energy and time
correlation analysis of the DAMA annual modulation data has been carried
out
\cite{Inel} handling
aspects other than the interaction type
as in ref. \cite{Sisd} (in this way a particular model framework is
fixed).

\begin{figure}[htb]
\centering
\vskip -0.4cm
\includegraphics[height=9.cm]{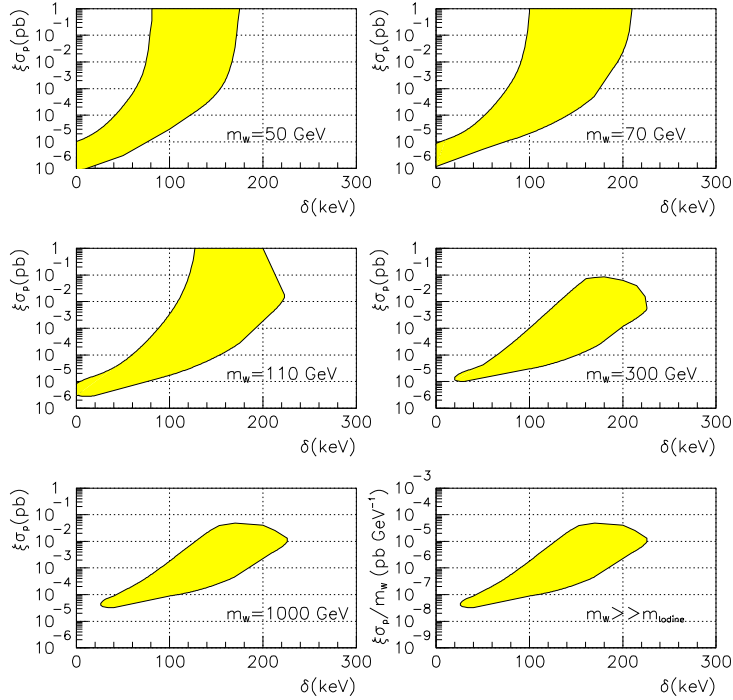}
\caption{An inelastic case: slices at fixed WIMP masses of the volume
allowed at 3 $\sigma$ C.L. in the space ($\xi \sigma_p$, $\delta$, $m_W$)
obtained for the model framework considered in ref. \cite{Inel}; some of
the uncertainties on used parameters have been included
\cite{Inel}. Note that e.g. Ge experiments
cannot explore most of the DAMA allowed regions in
this scenario.}
\label{fg:fig5}
\vskip -0.4cm
\end{figure}

In this scenario of Dark Matter with inelastic scattering an allowed
volume
in the space ($\xi \sigma_p$,$m_W$,$\delta$)
is obtained \cite{Inel}. For simplicity, Fig. \ref{fg:fig5} shows slices
of such an allowed volume at some given WIMP masses (3 $\sigma$ C.L.).
It can be noted that when $m_W \gg m_N$,
the expected differential energy spectrum is trivially dependent on $m_W$
and in particular it is proportional to the ratio between $\xi \sigma_p$
and $m_W$; this particular case is summarized in the last
plot of Fig. \ref{fg:fig5}.
The allowed regions have been obtained
-- as in the previous cases -- by the superposition
of those obtained when varying the values of the previously
mentioned parameters according to their uncertainties.
Note that
-- as in the previous cases --
each set of values (within those allowed by
the associated uncertainties) for the previously mentioned
parameters gives rise to a different expectation, thus to different
"most likely" values.
As an example we mention that when fixing
the other parameters as in Ref. \cite{Sisd},
the "most likely" values for a WIMP mass of 70 GeV
are: i) $\xi \sigma_p$ = $2.5 \times
10^{-2}$ pb and $\delta = 115$ keV when $v_0 = 170$ km/s,
ii) $\xi \sigma_p$ = $6.3 \times
10^{-4}$ pb and $\delta = 122$ keV when $v_0 = 220$ km/s; they are in
$\delta$ region were Ge and Si experiment are disfavoured.
Finally, we note also here that significant enlargement of the given
allowed regions
should be expected when including complete effects of model (and related
experimental and theoretical parameters) uncertainties.

\section{Proofs and Disproofs}

Let us preliminary remark that the claim for contradiction made by some
authors (see e.g. \cite{CDMS,CDMS2}) which use Ge target nuclei and
different
methodological approach is intrinsecally wrong.
In fact (besides the usual uncertainties which always exist in the
comparison of results achieved by different experiments):
i) the annual modulation signature gives a model independent
evidence for WIMPs independently on their nature and coupling, while an
exclusion plot is always model dependent;
thus, no direct and model-independent
comparison can be pursued among experiments which use different
approaches, different
techniques and, even more, different target nuclei; ii)
within the same general assumptions for a model e.g. SI coupling and
isothermal halo, the proper accounting for
parameters uncertainties, scaling laws uncertainties etc.  significantly
extends
the allowed region and moves the best fit values;
iii) there exist
many model frameworks to which Na and I are sensitive and
other nuclei (e.g. Ge and Si) are not. For example, a possible
WIMP with a SI cross section of few 10$^{-7}$ pb and with
SD cross section of few 10$^{-1}$ pb would produce a sizeable signal
in DAMA but almost nothing in Ge and Si experiments.
Moreover, also the given exclusion plots have a relative meaning
since they are calculated under many assumptions; thus,
they can drastically change either from model to model or
just changing the value of each used experimental and theoretical
parameter within its uncertainty.

Let us shortly remind that some comments about other
discussions of purely SD component
as well as about the
comparisons with other direct and indirect experiments
can be found in ref. \cite{Sisd}. Here, we only remind
that the HEAT balloon experiment has confirmed an excess
of high energy positrons in cosmic rays which - if interpreted in terms
of WIMP annihilation in a given particular scenario - gives rise to a
result compatible
with that of DAMA within the uncertainties
(see e.g. \cite{Kane}).

In the following only few comments on the CDMS-I result
and on its
comparison with that of DAMA are given.

As a general comment on the CDMS-I data \cite{CDMS} as well as on their
re-analysis in \cite{CDMS2},
the arguments of our ref. \cite{texas} still hold. In
addition, among the
many crucial items necessary to credit the result itself
and which are still missing, we cite:
i)  the measured energy spectra of gamma and electrons sources
    in running conditions. This can allow
    to show in a direct and safe way the
    detectors' response (from the peak positions and widths) and
    to derive the bulk detection efficiency;
ii)  the systematics associated to the used procedures
     since the measured counting rate is 60 counts keV$^{-1}$ kg$^{-1}$
d$^{-1}$
     (about 60 $\times$ 90keV $\times$ 15.8 kg $\times$ day = 85000
events,
     for the considered selected
     data sample) and only few of them (23 in the re-analysis)
	survive the veto,
     the so-called "quality cuts" (made - as mentioned in ref.
     \cite{CDMS2} - to
     remove large periods with hardware troubles) and
     the
     discrimination. In addition, these few events are mainly discarded by
     means of a neutron background modelling which suffers of some flaws,
     e.g.:
     a) the systematics due to the large amount of so-called "surface
events"
     present in all the
     regions of the discrimination plot for the Si data
	(as shown by CDMS so far and
     now not included in Fig. 40 of \cite{CDMS2});
     b) the uncertainties in the calculations based on the multiple
     scattering events. In fact, 4 events with both ionization yields Y1
     and Y2 $<$  0.5 (that is neutron candidates) and about 60-70 events
     with Y1,Y2 about 1 (that is gamma/electron candidates) have been
     counted. The small
     number of gamma/electron
     multiple scattering has to be compared with the number of
     gamma/electron single scattering (after veto
     reduction procedure about
     2 keV$^{-1}$ kg$^{-1}$ d$^{-1}$  $\times$ 90keV $\times$ 15.8 kg
$\times$ day =
     2800 events).
     This is a relevant point since the neutron background modelling is
     based
     on the comparison of events with Y$<$0.5: 4 multiple-scatterings
versus
     23 single-scatterings.

Furthermore, let us quickly comment about the comparison that generally is
done
- see e.g. \cite{CDMS2} - with the DAMA result.

As mentioned above in many models Ge and Si target nuclei are insensitive
to
WIMP candidate to which instead Na and I are. Furthermore, in the models
where all of them are sensitive,
in order to obtain cross-section limits/regions from the experimental
data, it is necessary to assume
an astrophysical, particle and nuclear Physics model framework, to
define a set of parameters values and to scale the cross sections
on the different
target nuclei to the one on proton.
Therefore, every inferred result
strongly depends e.g. on the chosen model, on the set of values fixed for
the
experimental and theoretical parameters and on the scaling laws
for each involved nucleus. Thus, large uncertainty is
associated to the results. This is a crucial point that must be
accounted every time a comparison among results
from different experiments and with theoretical models is attempted,
avoiding that a reader might have the wrong idea of the
"universality" of the conclusion.

\begin{figure}[htb]
\centering
\vskip -0.6cm
\includegraphics[height=6.4cm]{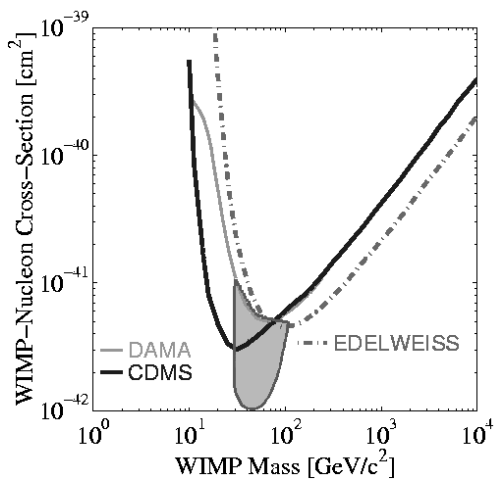}
\caption{A model dependent comparison for purely SI coupled WIMP:
exclusion limits claimed by
CDMS \cite{CDMS2} and EDELWEISS \cite{Edel}, superimposed to the DAMA
region
allowed in the simplified model with isothermal
galactic halo, "simple" scaling laws and fixed assumptions for all the
parameters; in particular, here, for the comparison $v_0$ has been
set to the value 220 km/s (that the
effect of its existing uncertainty
on the allowed region - correctly accounted in ref. \cite{Mod3} and here
in the previous
fig. \ref{fg:fig2} - has been neglected). This figure
is as in ref. \cite{CDMS2}, but here the
correct region allowed by the cumulative DAMA data under the mentioned
assumptions
is shown. We further remind that the proper inclusion of the existing
models' uncertainties
significantly enlarges the allowed region (see e.g. the case of inclusion
of $v_0$ uncertainties and
of halo modelling in Fig. \ref{fg:fig2} and \ref{fg:fig3}, respectively).
Also exclusion plots do not show "universal" boundary, but are affected
by large
experimental and theoretical uncertainties.}
\label{fg:fig6}
\end{figure}

In particular, as regards the purely spin-independent coupled WIMPs,
which is practically the only scenario considered by other authors,
the comparisons are often arbitrarily performed not with the region
correctly endorsed by
the DAMA collaboration (that is
including the constraint from the measured upper limits)
for the given simplified
model framework of ref. \cite{Mod3}
(see again  \cite{CDMS2}). In fact,
as an example, in Fig. \ref{fg:fig6} the exclusion limits claimed
by CDMS and EDELWEISS are superimposed to the correct DAMA region
for the simplified model with isothermal
galactic halo, "simple" scaling laws and fixed assumptions for all the
parameters as in ref. \cite{Mod3}, but assuming here
the $v_0$ value at 220 km/s (that is neglecting the effect of its
existing uncertainty
on the allowed region, as correctly accounted in ref. \cite{Mod3} and here
in the previous
fig. \ref{fg:fig2}).

Moreover, in ref. \cite{CDMS2} a comparison is pursued by using
some "most likely" values quoted in ref. \cite{Mod3};
however, it is worth to note
that these "most likely" values have not a general validity, but depend
on the specific assumptions (on halo, on particle, on nuclear physics) and
experimental (e.g. quenching factors, detector response, etc.) and
theoretical
(form factors, scaling laws, etc.) parameters' values.
Varying them within the existing  uncertainties
varies the "most likely" values.
Realistic estimates of the expected number of counts in CDMS-I from
the DAMA results
for the purely SI coupled WIMPs can vary
from about 20 down to about 1.
Moreover, let us remind that - even under the unrealistic
assumption of no systematics in the hardware and software CDMS
data reduction -
the calculated exclusion plot has not an "universal" validity.

A comparison between the Ge result and a particular model dependent
extrapolation from the DAMA 2-6 keV cumulative residuals has
also been addressed in ref.
\cite{CDMS2}; this is again model dependent and, in particular, depends
on the used quenching factors, form factor parameters, etc.
In particular, the used procedures does take into account neither the
energy behaviour of the measured modulation amplitude and the upper
limits on recoils nor the proper values and uncertainties
of parameters needed in the calculation.  Thus, considerations similar to
the previous case hold.

Finally, in Fig. 47 of ref. \cite{CDMS2} --
as sometimes happens in other presentations --
the theoretical expectations for the particular case of the neutralino
are shown without taking into account the uncertainties on the
local halo density and the rescaling procedures (as well as on some other
astrophysical, nuclear and particle physics aspects). This incorrect
procedure leads to wrong conclusion.

Summarizing, there is no solid scientific reason for the CDMS claim
for contradiction whatever scenario would be considered.

\section{Towards LIBRA}

At present our main efforts are devoted to the realization
of LIBRA (Large sodium Iodine Bulk for RAre processes in the DAMA
experiment) consisting of $\simeq$ 250 kg of NaI(Tl). New radiopure
detectors by chemical/physical purification of NaI and Tl powders as a
result of a dedicated R$\&$D with Crismatec have been realized.
The installation of this set-up is foreseen in fall 2002;
this will allow us to increase the sensitivity of
the experiment. Some related arguments can be found in
ref. \cite{erice}.

\section{Conclusions}

The DAMA annual modulation data of four annual cycles
\cite{Mod1,Mod2,Ext,Mod3,Sist,Sisd,Inel}
have been analysed by energy and time correlation analysis
in terms of purely SI, purely SD, mixed SI/SD,
``preferred'' inelastic WIMP scattering model frameworks.

To effectively discriminate among the different possible scenarios
further investigations are in progress.
In particular, the data of the 5$^{th}$ and
6$^{th}$ annual cycles are at hand, while the set-up is running to collect
the data of a 7$^{th}$ annual cycle. Moreover, the LIBRA (Large sodium
Iodine Bulk for RAre processes) set-up is under
construction to increase the experimental sensitivity.

\end{document}